\documentclass[11pt]{article}
\usepackage{graphicx}
\usepackage{amsmath,amssymb}
\def\1{{\bf 1}}

\def\be{\begin{equation}}
\def\ee{\end{equation}}

\begin{document}
\begin{center}
{\Large\bf  A Brief Review on Results and Computational Algorithms for Minimizing the Lennard-Jones Potential}
\end{center}

\bigskip

\begin{center}
{\Large\bf Jiapu Zhang}

\bigskip

{\it Centre for Informatics and Applied Optimization \&\\
Graduate School of Information Technology and Mathematical Sciences,\\
University of Ballarat, Victoria 3353, Australia}
\end{center}

\bigskip

\begin{abstract}
The Lennard-Jones (LJ) Potential Energy Problem is to construct the most stable form of
$N$ atoms of a molecule with the minimal LJ potential energy.
This problem has a simple mathematical form
\begin{equation*}
\mbox{minimize}~~~f(x)=4\sum_{i=1}^N \sum_{j=1,j<i}^N \left(
\frac{1}{\tau_{ij}^6} -\frac{1}{\tau_{ij}^3}
\right)~~~\mbox{subject to}~~~ x\in \mathbb{R}^n,
\end{equation*}
\noindent where $\tau_{ij}=(x_{3i-2}-x_{3j-2})^2
                +(x_{3i-1}-x_{3j-1})^2
                +(x_{3i}  -x_{3j}  )^2$,
$(x_{3i-2},x_{3i-1},x_{3i})$ is the
coordinates of atom $i$ in $\mathbb{R}^3$,
$i,j=1,2,\dots,N(\geq 2 \quad \text{integer})$, and $n=3N$;
however it is a challenging and difficult problem for many
optimization methods when $N$ is larger. In this paper, a
brief review and a bibliography of important computational algorithms on minimizing the LJ potential energy are introduced in Sections 1 and 2. Section 3 of this paper illuminates many beautiful graphs (gotten by the author nearly 10 years ago) for the three dimensional structures of molecules with minimal LJ potential.  
\end{abstract}

\vskip 0.2cm

\section{Introduction}
the Lennard-Jones potential energy problem is
to construct the most stable form of
$N$ atoms of some material with the minimal energy structure.
Its form in mathematics is very simple:
\begin{equation}
\mbox{minimize}~~~f(x)=4\sum_{i=1}^N \sum_{j=1,j<i}^N \left(
\frac{1}{\tau_{ij}^6} -\frac{1}{\tau_{ij}^3}
\right)~~~\mbox{subject to}~~~ x\in \mathbb{R}^n, \label{ljproblem}
\end{equation}
\noindent where $\tau_{ij}=(x_{3i-2}-x_{3j-2})^2
                +(x_{3i-1}-x_{3j-1})^2
                +(x_{3i}  -x_{3j}  )^2$,
$(x_{3i-2},x_{3i-1},x_{3i})\in \mathbb{R}^3$ is the
coordinates of atom $i$,
$i,j=1,2,\dots,N(\geq 2 \quad \text{integer})$, and $n=3N$.
However, it interests many researchers in the field of
biology, physics, chemistry, mathematical optimization,
computer science, and materials science.
The reason lies in the nonconvexity of the objective function and the huge
number of local minima, which is growing exponentially with $N$.
\cite{hoare} tells us the number of distinct local minima of
$N$-atoms Lennard-Jones problem is about $O(e^{N^2})$.
For example, When $N=$6, 7, 8, 9, 10, 11, 12, 13 the numbers of distinct
local minima of the problem are 2, 4, 8, 18, 57, 145, 366, 988
respectively (\cite{hoarem}).
From the point of view of numerical optimization methods, this problem is an
excellent test problem for local or global optimization methods.
However, the Lennard-Jones Problem is very difficult
and challenging to many optimization methods even
when $N$ is not very large. By now, the best objective function vlues for
$2\leq N\leq 309$ \cite{web1, web2, web3} are well known.
At first, around 1972 Hoare and Pal
\cite{hoare1, hoare2, hoare3} gave the best values for atoms
5$\sim$16, 18$\sim$21, 25, 26, 29, 55.
Then, in 1987, Northby (\cite{northby}) made a landmark$--$yielding
most of the lowest values in the range $13\leq N\leq 150$;
and in 1999 \cite{romerobg} presented
the best results on $151\leq N\leq 309$ with the exception of
$N=$192, 201 \cite{walesd}, $N$4=200, 300 \cite{xue},
$N=$185 \cite{learybob}, $N=$186, 187 \cite{hartke}.
At present, about thirty percent of Northby's results have been improved
\cite{xue, doye1, doye2, leary1, freeman,
farges, wille, coleman, deaven, wolf}.
The improvements are described as follows.
In 1994 Xue (\cite{xue}) got the best values for
$N=$65,66,134.
Doye et al (\cite{doye1,doye2}) in 1995 gave the best values on
$N$=38,75$\sim$77,102$\sim$104; and
Leary in 1997 (\cite{leary1}) presented best values for
$N=$69,78,88,107,113,115.
Wolf (\cite{wolf}) got the best values on
$N=$82, 84, 86, 88, 92, 93, 94, 95, 96, 99, 100 in 1998.
For the left so-called magic numbers sequence:
17,23,24,72,88, Freeman, Farges, Wille, Coleman, Deaven
(\cite{freeman, farges, wille, coleman, deaven}) got the best results, respectively.

The outline of this paper is as follows. In section 2, we review the
methods in the most important references above-mentioned. Section 3
contains a description of our techniques and methods to tackle
problem (\ref{ljproblem}); and results of numerical experiments for
the problem are given in this section. The structure of our optimal
solutions is illustrated, and can be compared with those of others,
at the end of this section. Section 4 concludes the paper.

\section{Approaches to the Lennard-Jones Problem}
Some good summaries on the methods solving the Lennard-Jones problem
in fact can be found in \cite{northby, doyewm, pardalossx, gockenbachks, leary1}.
In this section we review the methods in the most important references
\cite{hoare1, hoare2, hoare3, northby, romerobg,
walesd, xue, learybob, hartke, doye1, doye2, leary1, wolf,
freeman, farges, wille, coleman, deaven}.

Hoare and Pal's work (\cite{hoare1, hoare2, hoare3})
may be the early most successful results on Lennard-Jones problem.
The idea is using build-up technique to construct the initial solutions
which are expected to represent low energy states, and using those
initial solutions as starting points for
a local search method to relax to the optimal solution (\cite{hoare3}).
The starting seed is the regular unit tetrahedron with atoms at the
vertices, the obvious global optimal solution for $N=4$.
Beginning with this tetrahedron, Hoare and Pal added one atom at a time to
construct a sequence of polytetrahedral structures and at last
got good results up to $N=66$.
For example, for $N=5$ its
globally optimal trigonal bi-pyramid (bi-tetrahedron)
structure is gotten by adding an atom
at the tetrahedral capping position over a triangular face;
following the bi-tetrahedron structure, the optimal structure of $N=6$
is tri-tetrahedron (another known optimal structure for $N=6$ is
octahedron (using tetrahedral capping over triangular faces and
half-octahedral capping over square faces),
which is not a polytetrahedron);
for $N=7$ its best structure constructed is the pentagonal bi-pyramid,
a structure with a five-fold axis of symmetry. Many computer science
data structure procedures such as greedy forward growth operator and reverse
greedy operator can make the build-up technique work well.
\cite{hoare1} describes the application of methods of studying noncrystalline
clusters to the study of ``spherical"
face centred cubic (fcc) microcrystallites.
\cite{hoare2} describes the chief geometrical features of the clustering of
small numbers of interacting particles.

The data structure of Northby (\cite{northby})
in finding the good starting solution
is the lattice based structure. The lattice structures consist of
an icosahedral core and particular combinations of surface lattice points.
\cite{mackay} first constructed a class of icosahedral packings by
adding successively larger icosahedral shells in layers around a core
central atom; this icosahedral lattice can be described as 20
slightly flattened tetrahedrally shaped fcc units with 12
vertices on a sphere centered at the core atom. Atoms within each
triangular face are placed in staggered rows in a two dimensional
hexagonal close-packed arrangement. Each atom in the interior of
a face in a given shell is a tetrahedral capping position
relative to three atoms in the underlying shell. Northby relaxed the
structure of \cite{mackay} to get his
IC and FC multilayer
icosahedral lattice structures.
The IC lattice can be referred to the FORTRAN code in \cite{xue};
it consists of all those sites which will comprise the outer shell of
the next complete Mackay (\cite{mackay}) icosahedron.
FC lattice is a slight modification of IC lattice in that its outer shell
maintains icosahedral symmetry and consists of points at the
icosahedral vertices and the stacking fault positions of the outer IC shell.
Basing on the IC and FC lattices, Northy gave his algorithm first finding
a set of lattice local minimizers and then relaxing those lattice minimizers by
performing continuous minimization starting with those lattice minimizers.
The algorithm was summarized as Algorithm 1 and Algorithm 2 of \cite{xue}.

The great majority of the best known solutions of Northy are icosahedral
in character. The hybridization of global search and local search methods,
usually, is more effective to solve the large scale problem
than the global search method or local search method
working alone. Catching those two ideas,
\cite{romerobg} combined a genetic algorithm with
a stochastic search procedure on icosahedrally derived lattices.
The structures of the optimal solutions gotten in \cite{romerobg} are
either icosahedral or decahedral in character.
The best results of \cite{wolf}
for $N=82,84,86,88,92,93,94,95,96,99,100$
were gotten by using a genetic algorithm alone.
Deaven et al (\cite{deaven}) also using the genetic algorithm got the
optimal value known for the magic number $N=88$.

The successful works to improve Northby's results
(\cite{northby}) were mainly done by
Xue (\cite{xue}), Leary (\cite{leary1}),
and Doye et al (\cite{doye1,doye2}).

Xue (\cite{xue}) introduced a modified version of the Northby algorithm.
He showed that in some cases the relaxation of the outer shell lattice local
minimizer with a worse potential function value may lead to a local
minimizer with a better value. In Northby's algorithm
the lattice search part is a discrete optimization local search procedure,
which makes a lattice move to its neighboring lattice with $O(N^{\frac{5}{3}})$
time complexity. In \cite{xue1} Xue introduced a simple storage data structure
to reduce the time complexity to $O(N^{\frac{2}{3}})$ per move; and then
used a two-level simulated annealing algorithm within the supercomputer CM-5
to be able to solve fastly the Lennard-Jones problem with
sizes as large as 100,000 atoms. In \cite{xue}
by employing AVL trees (\cite{horowitzs}) data
structure Xue furthermore reduced the time complexity to $O(\log N)$ if NN
(nearest neighbor) potential function is used. He (\cite{xue}) relaxed
{\it every} lattice local minimizer found instead of relaxing only those
lattice local minimizers with best known potential function value by a
powerful Truncated Newton local search method,
and at last got the best results known for
$N=65,66,134,200,300$.

Leary (\cite{leary1}) gave a successful Big Bang Algorithm in 1997 for getting
the best values known of $N=69,78,88,107,113,115$.
In \cite{leary1} the FCC lattice structure is discussed and its connections
are made with the macrocluster problem. It is also concluded in \cite{leary1}
that almost all known exceptions to global optimality of the well-known
Northby multilayer icosahedral conformations for microclusters are shown to be
minor variants of that geometry. The Big Bang Algorithm contains 3 steps:
Step 1 is an initial solution generating procedure which randomly generates
each coordinate of the initial solution with
the independently normal distribution;
Step 2 is to generate the new neighborhood solution by
discrete-typed fixed step steepest descent method, which is
repeated until no further progress is made;
Step 3 is to relax the best solution gotten in Step 2 by a
continuous optimization method$--$conjugate gradient method.

Doye et al (\cite{doye1}) investigated the structures of clusters by
mapping the structure of the global minimum as a function of both cluster
size and the range of the pair potential which is appropriate to the clusters of
diatomic molecule, $C_{60}$ molecule, and the ones between them both.
For the larger clusters the structure of the global minimum changes
from icosahedral to decahedral to fcc as the range is decreased (\cite{doye1}).
In \cite{doye2} Doye et al predicted the growth sequences for small
decahedral and fcc clusters by maximisation of the number of
NN contacts.

Lastly, in this section, we give a brief review of methods on the magic numbers
$N=17,23,24,72,88$ and on the exceptions to \cite{romerobg}.
Freeman et al (\cite{freeman})
presented the best value for $N=17$ when the
thermodynamic properties of argon clusters were studied by a combination
of classical and quantum Monte Carlo methods.
The polyicosahedral growth of Farges et al (\cite{farges})
starts from a 13-atom primitive icosahedron
containing a central atom and 12 surface
atoms. On each one of the five tetrahedral sites, surrounding
a particular vertex, a new atom is added and finally a sixth atom
is placed on top to create a pentagonal cap. In this way
a 19-atom structure being made of double interpenetrating
icosahedra, which is a 13-atom icosahedra sharing
9 atoms, is obtained. I.e., for three pentagonal bipyramids
each one shares an apex with its nearest neighbour.
In this way a 23-atom model
consisting of three interpenetrating icosahedra is gotten for the best
value known. Wille (\cite{wille}) used the simulated annealing method
yielding low-lying energy states whose distribution depends on the
cooling rate to find the best solution known for $N=24$.
Coleman et al (\cite{coleman}) proposed a build-up process
to construct the optimal solution structures.
The HOC (half octahedral cap) structure of the optimal solution for $N=72$
is found by a prototype algorithm
designed using the anisotropic effective energy simualted
annealing method at each build-up stage (\cite{coleman}).
Wales and Doye (\cite{walesd}) in 1997 gave the lowest values known for
$N=192,201$. Their method is so-called basin-hopping method, in which
first the transformed function $\tilde{f}(x)=\min \{ f(x)\}$ was defined
and performed starting from $x$ by the
PR conjugate gradient method (see, for example, \cite{sunz}),
and then the energy landscape for the function $\tilde{f}(x)$ was
explored using a canonical Monte Carlo simulation.
Leary (\cite{learybob}) has developed techniques for moving along sequences
of local minima with decreasing energies to arrive at good candidates
for global optima and got the best value known on $N=185$.

\section{Our approach to the Lennard-Jones Problem}
In this section we first briefly introduce
the discrete gradient method; then present
a hybrid discrete gradient and simulated annealing method. And then
we give some techniques, with the discrete gradient method
and the hybrid method, to tackle the Lennard-Jones Problem.

\subsection{Discrete Gradient Method}
Discrete gradient method is a
derivative-free local search method for nonsmooth optimization
(\cite{bagirov,bagirov1}). Results of numerical experiments
presented in \cite{bagirovr2} show that this method can jump over
stationary points, which are not local minima, so we can reduce
the number of stationary points, which we meet. Details of this
method can be found, for example, in \cite{bagirov,bagirov1}.

\subsection{Hybrid Discrete Gradient and Simulated Annealing Method}
Simulated annealing method is a global search method. It has
received a great deal of attention in last several years. First,
this method was applied to combinatorial optimization, i.e. when
the objective function is defined in a discrete domain (see
\cite{cerny85, kirkpatrick83}). Later this method was applied to
solve continuous global optimization problems (see \cite{brooks88,
corana87, jones95, locatelli00, romeijn94}). Convergence of
simulated annealing method is studied, for example, in
\cite{locatelli00}.

Numerical experiments show that algorithms based on a combination
of the global and local search techniques are more effective than
the global search methods working alone (see, for example,
\cite{bagirovr2,yiult,hedarf}). In these methods a local search
algorithm is used to find a stationary point and a global search
algorithm is used to escape from this stationary point and to find
a new starting point for a local search algorithm.
In this subsection, we develop a new hybrid simulated annealing and
discrete gradient method. During the whole
process of the method, both simulated annealing and discrete
gradient methods need the the objective function values only.

The simulated annealing method for solving continuous global
optimization problems was studied by many authors (see, for
example, \cite{brooks88, corana87, jones95, locatelli00,
romeijn94}). {\it Neighborhood solution search} procedure is one
of the important steps in this method. Noticing that the
neighborhood solution search for simulated annealing method should
be at least based on two basic ideas: (a) neighbor means
``nearby", (b) simulated annealing method is a stochastic method
so that the neighborhood solution should be randomly taken, we may
simply give a neighborhood solution search procedure for simulated
annealing algorithm: {\small {\sf
\begin{enumerate}
\item[] Uniformly randomly keeping $n-1$ coordinates of $x$, and making the left
one coordinate of $x$ randomly take a value such that the new
solution $x'$ still feasible. This gives $x'$.
\end{enumerate}
} } \noindent When the feasible region of the optimization problem
is the unit simplex $S=\{x\in \mathbb{R}_+^n :\sum_{i=1}^n
x_i=1\}$, the neighborhood solution search procedure should be
done a little modification: {\small {\sf
\begin{enumerate}
\item[] Uniformly randomly keeping $n-2$ coordinates of $x$, and making
one coordinate from the two elements left to $x$ randomly take a
value from [0,1] such that the value of the sum of the $n-1$
coordinates gotten not greater than 1. Another left coordinate of
$x'$ is given the value 1-sum. This gives $x'$.
\end{enumerate}
}}

The hybrid method starts from an initial point, first executes
discrete gradient method to find local minimum, then carries on
simulated annealing method in order to escape from this local
minimum and to find a new starting point for the discrete gradient
method. Then we again apply the discrete gradient method starting
from the last point and so on until the sequence of the optimal
objective function values gotten is convergent. The pseudo-code of
the hybrid method is listed as following:

\vskip 0.3cm

\noindent {\sf Algorithm 1: Hybrid simulated annealing and
discrete gradient method}\\
{\small
{\sf
{\bf \underline{Initialization:}}
\begin{enumerate}
\item[] {\it Define the objective function $f$ and its feasible solution space}.
\item[] Call the {\it initial feasible solution generating procedure} to get $x$.
\item[] Call {\it initial temperature selecting procedure} to get $T$.
\item[] Inialization of $f$: $f=f(x)$.
\item[] Initialize the neigbourhood feasible solution: $x\_{neighbour}=0$.
\item[] Initialize $x\_{best}$: $x\_{best}=x$.
\item[] Initialize $f\_{best}$: $f\_{best}=f$.
\end{enumerate}
do $\{$
\begin{enumerate}
\item[] {\bf \underline{Discrete Gradient local search part:}}
\item[] $\qquad f\_{best}\_{local} = {\bf local\_{search}} (x\_{best},
x\_{new}\_{gotten})$;
\item[] $\qquad x=x\_{new}\_{gotten}$;
\item[] {\bf \underline{Simulated Annealing global search part:}}
\begin{enumerate}
\item[] do $\{$
\item[] $\qquad$ do $\{$
           \begin{enumerate}
           \item[] $\qquad$ $x\_{neighbour}=randomly\_{perturb}(x)$;
           \item[] $\qquad$ $f\_{neighbour}=f(x\_{neighbour})$;
           \item[] $\qquad $ Calculate the difference $\Delta=f\_{neighbour}-f$;
           \item[] $\qquad$ {\bf If} ($\Delta \leq 0$) {\bf or}
                   (random[0,1] $<$ exp(-$\Delta /T$))
           \item[] $\qquad \qquad x=x\_{neighbour} \quad f=f\_{neighbour}$;
           \item[] $\qquad$ {\bf If} ($f \leq f\_{best})$ $\quad$
                   $\quad x\_{best}=x \quad f\_{best}=f$;
           \end{enumerate}
\item[] $\qquad$ $\}$ while (equilibrium has not been reached);
\item[] $\qquad$ Temperature annealing
\item[] $\}$ while (Simulated Annealing stop criterion has not been met);
\end{enumerate}
\end{enumerate}
$\}$ while ( $f\_{best} - f\_{best}\_{local} \leq -0.001$ );
}
}
\vskip 0.4cm

The convergence of the proposed hybrid method directly follows
from the convergence of the simulated annealing and the discrete
gradient methods.

The simulated annealing method part consists of two procedures: inner
procedure for the search of neighborhood solution and outer
procedure for the cooling temperature $T$. The number of
iterations of inner and outer loops are taken to be large enough
for guaranteeing that the sufficient iterations will be carried on
to escape from the current local minimum.
In implementing the hybrid method, we use $T=0.9*T$ as the
temperature annealing schedule and the initial temperature is
taken large enough according to the rule in \cite{kirkpatrick83}.
We restrict the number of iterations for the outer procedure by
100 and number of iterations for the inner procedure by 1000. The
discrete gradient method part is terminated when the distance
between the approximation to the subdifferential and origin is
less than a given tolerance $\epsilon > 0$ ($\epsilon =10^{-4}$).

\subsection{Minimization of Lennard-Jones potential function}\label{protein}
In this subsection, the proposed discrete gradient method
and hybrid discrete gradient and simulated annealing method,
with the techniques given in the below,
have been applied for minimization of
well-known Lennard-Jones potential function in protein folding
problem (\ref{ljproblem}).
The nonconvex Lennard-Jones potential function
has a huge number of local minima. Therefore many global
optimization techniques fail to minimize it. The proposed hybrid
method fails to solve this problem when number of atoms $N \geq
20$. In order to reduce the number of local minima we suggest to
approximate the function
$$
\varphi(\tau) = \frac{1}{\tau^6} -\frac{1}{\tau^3}
$$
by the following function:
$$
g(\tau)=\max (g_1(\tau),\min (g_2(\tau), g_3(\tau)))
$$
where $g_1(\tau)$ is the piecewise linear approximation of the
function $\varphi(\tau)$ in segment $(0,r_0]$, $g_2(\tau)$ is the
piecewise linear approximation of this function over segment
$[r_0,r_1]$, and finally $g_3(\tau)$ is the piecewise linear
approximation over $[r_1,b]$ and $b$ is large enough number. Here
$$
r_0=\sqrt[3]{2},~~r_1=1/\sqrt[3]{2/7}
$$

Such a approximation of the function $\varphi(\tau)$ allows us to
remove many local minima of the Lennard-Jones potential function
and to get a good approximation to the global minimum of the
objective function $f$ in problem (\ref{ljproblem}).

In numerical experiments we take $b=16$ and divide the segment
$[0.001,r_0]$ into 100 segments, the segment $[r_0,r_1]$ into 100
segments and the $[r_1,16]$ into 50 segments which allows one to
get good approximations for the function $\varphi(\tau)$.
The replacement of the function $\varphi(\tau)$ by the function
$g(\tau)$ makes the objective function nonsmooth. On the other
side such a replacement significantly reduce the number of local
minima. Since the discrete gradient method is a method of
nonsmooth optimization the proposed hybrid method can be applied
for solving this transformed problem.

When solving the Lennard-Jones problem,
first we use the discrete gradient method with build-up
technique to relax to an initial solution. Then we apply the hybrid
discrete gradient and simulated annealing method, with the good
approximation for the objective function, to get another initial solution.
Starting from this initial solution we again apply the derivative-free
discrete gradient method and at last get the global solution.
Results of numerical experiments are presented in Table
\ref{table1} as the appendix. Results from Table
\ref{table1} show that our techniques and methods effectively solve protein
folding problem when number of atoms is not greater than 310.

\begin{table}[h!]
\caption{Our numerical results for Lennard-Jones Protein Problem}
\begin{center}
\begin{tabular}{|c|c|c|}\hline
Number of atoms   &Best value obtained  &Best value known\\ \hline
19        &-72.659782        &-72.659782\\ \hline
20        &-77.177043        &-77.177043\\ \hline
21        &-81.684571        &-81.684571\\ \hline
22        &-86.573675        &-86.809782\\ \hline
23        &-92.844461        &-92.844472\\ \hline
24        &-97.348815        &-97.348815\\ \hline
25        &-102.372663       &-102.372663\\ \hline
27        &-112.825517       &-112.873584\\ \hline
30        &-128.096960       &-128.286571\\ \hline
34        &-150.044528       &-150.044528\\ \hline
44        &-207.631655       &-207.688728\\ \hline
49        &-239.091863       &-239.091864\\ \hline
56        &-283.324945       &-283.643105\\ \hline
65        &-334.014007       &-334.971532\\ \hline
67        &-347.053308       &-347.252007\\ \hline
84        &-452.267210       &-452.6573\\ \hline
93        &-510.653123       &-510.8779\\ \hline
148       &-881.072948       &-881.072971\\ \hline
170       &-1024.791771      &-1024.791797\\ \hline
172       &-1039.154878      &-1039.154907\\ \hline
268       &-1706.182547      &-1706.182605\\ \hline
288       &-1850.010789      &-1850.010842\\ \hline
293       &-1888.427022      &-1888.427400\\ \hline
298       &-1927.638727      &-1927.638785\\ \hline
300       &-1942.106181      &-1942.106775\\ \hline
301       &-1949.340973      &-1949.341015\\ \hline
304       &-1971.044089      &-1971.044144\\ \hline
308       &-1999.983235      &-1999.983300\\ \hline
\end{tabular}
\end{center}
\label{table1}
\end{table}

The structure of the optimal solutions corresponding to the above
optimal values is illustrated and can be compared with the ones of
\cite{web1,romerobg} in the following figures:

\begin{figure}
 \label{ouratoms}
 \centering
\includegraphics[scale=0.4]{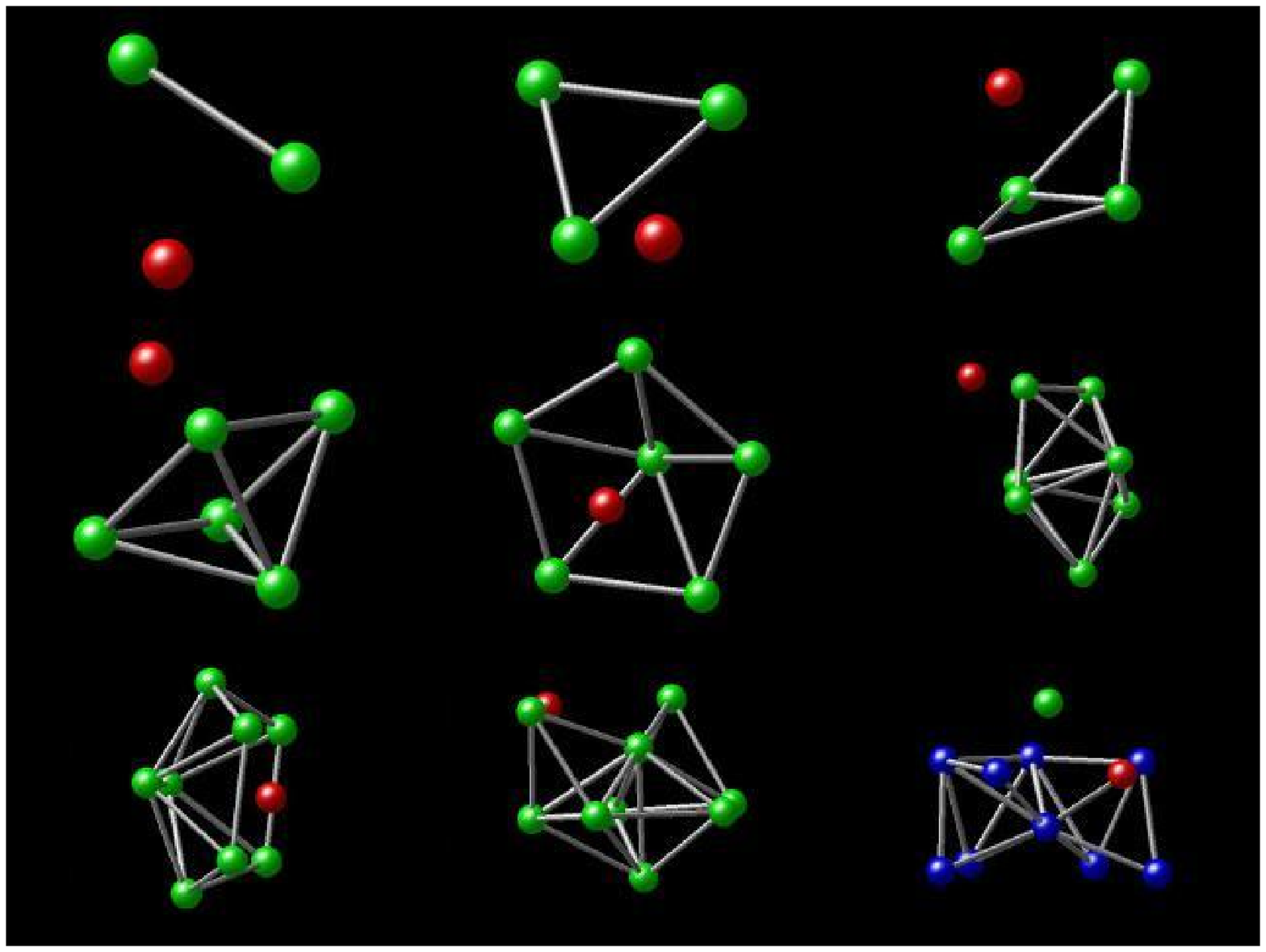}
\includegraphics[scale=0.4]{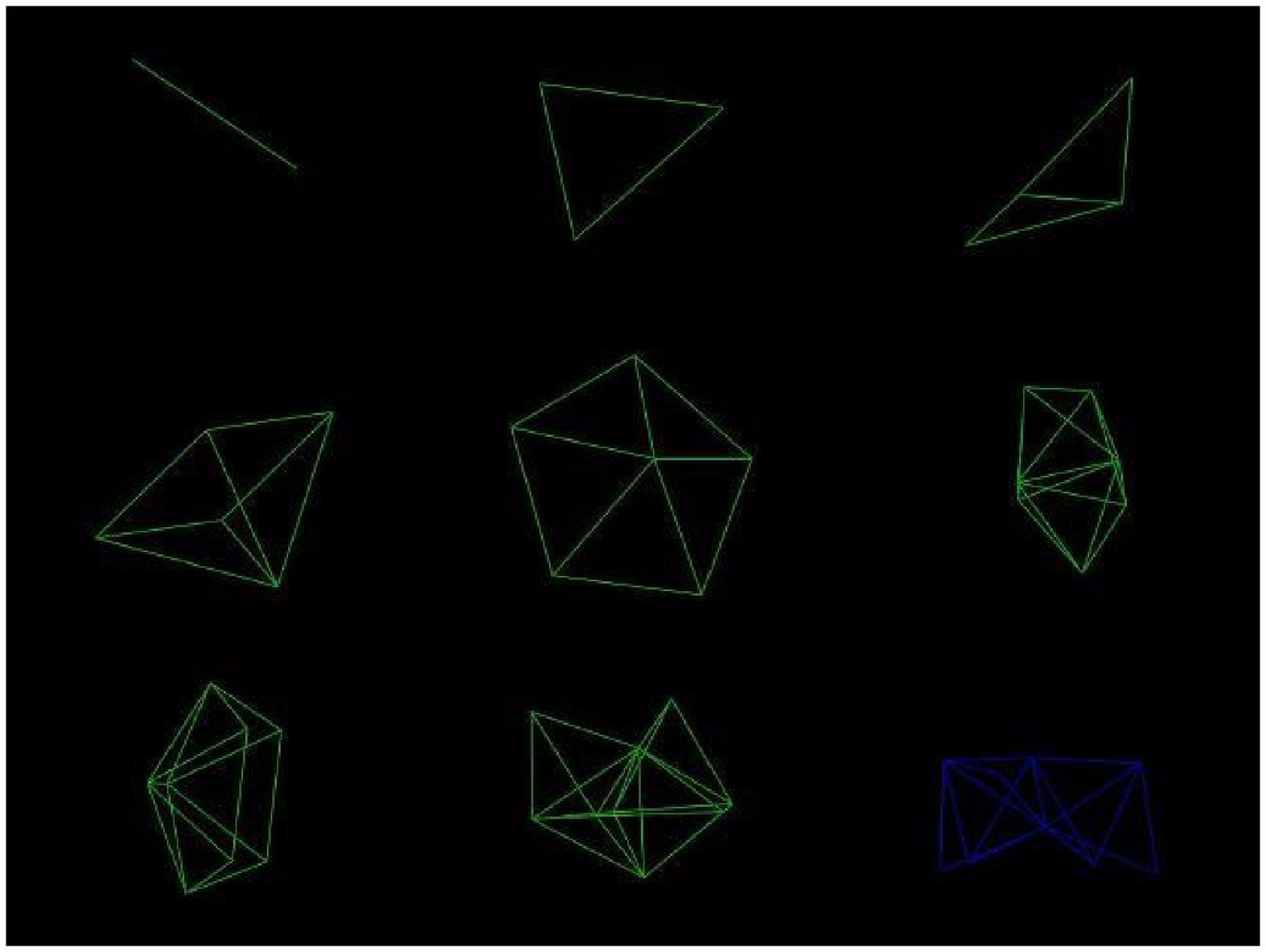}
\text{N=3, 4, 5, 6, 7, 8, 9, 10, 11 (ours)}
\end{figure}

\begin{figure}
 \label{ouratoms}
 \centering
\includegraphics[scale=0.4]{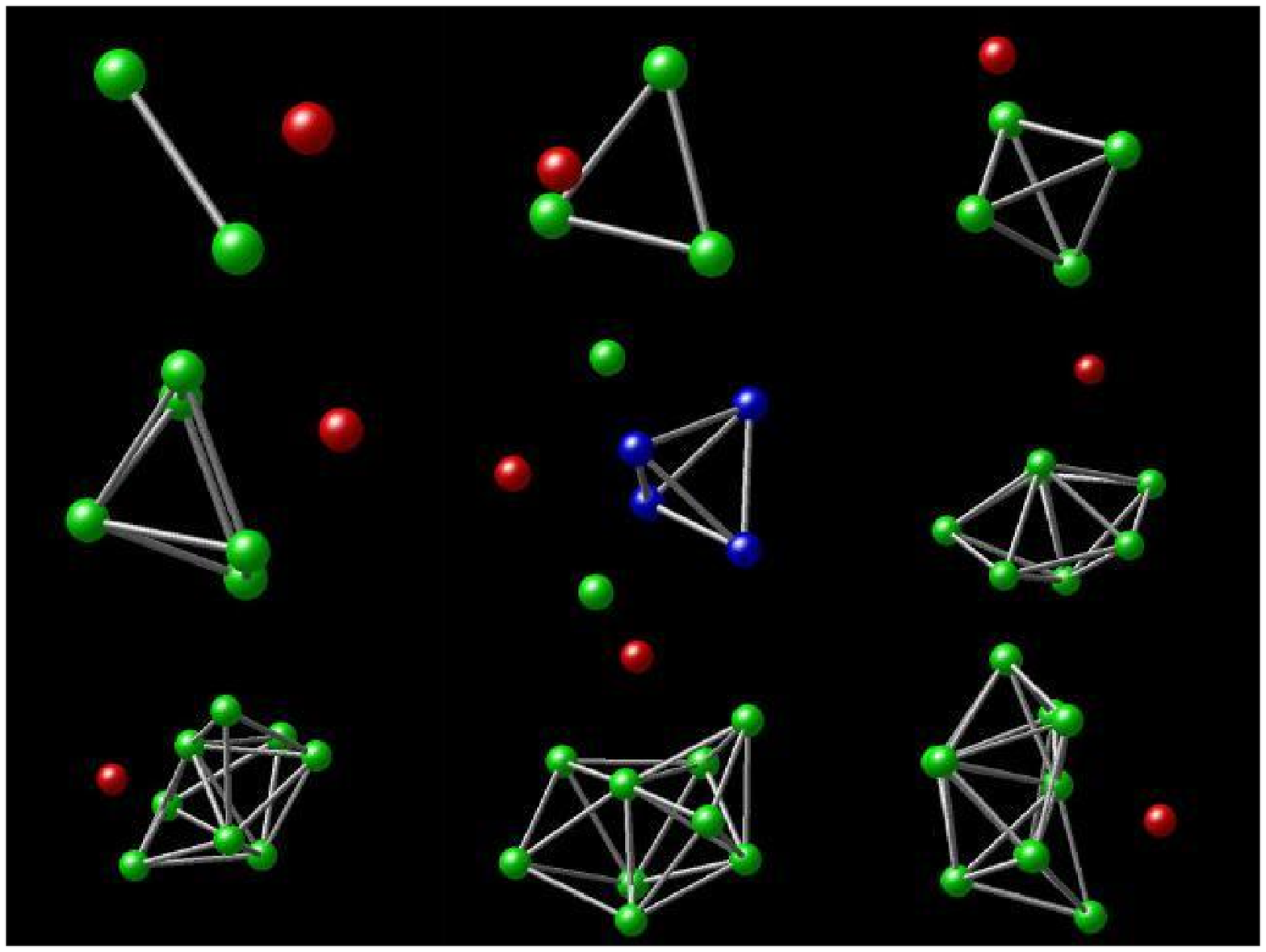}
\includegraphics[scale=0.4]{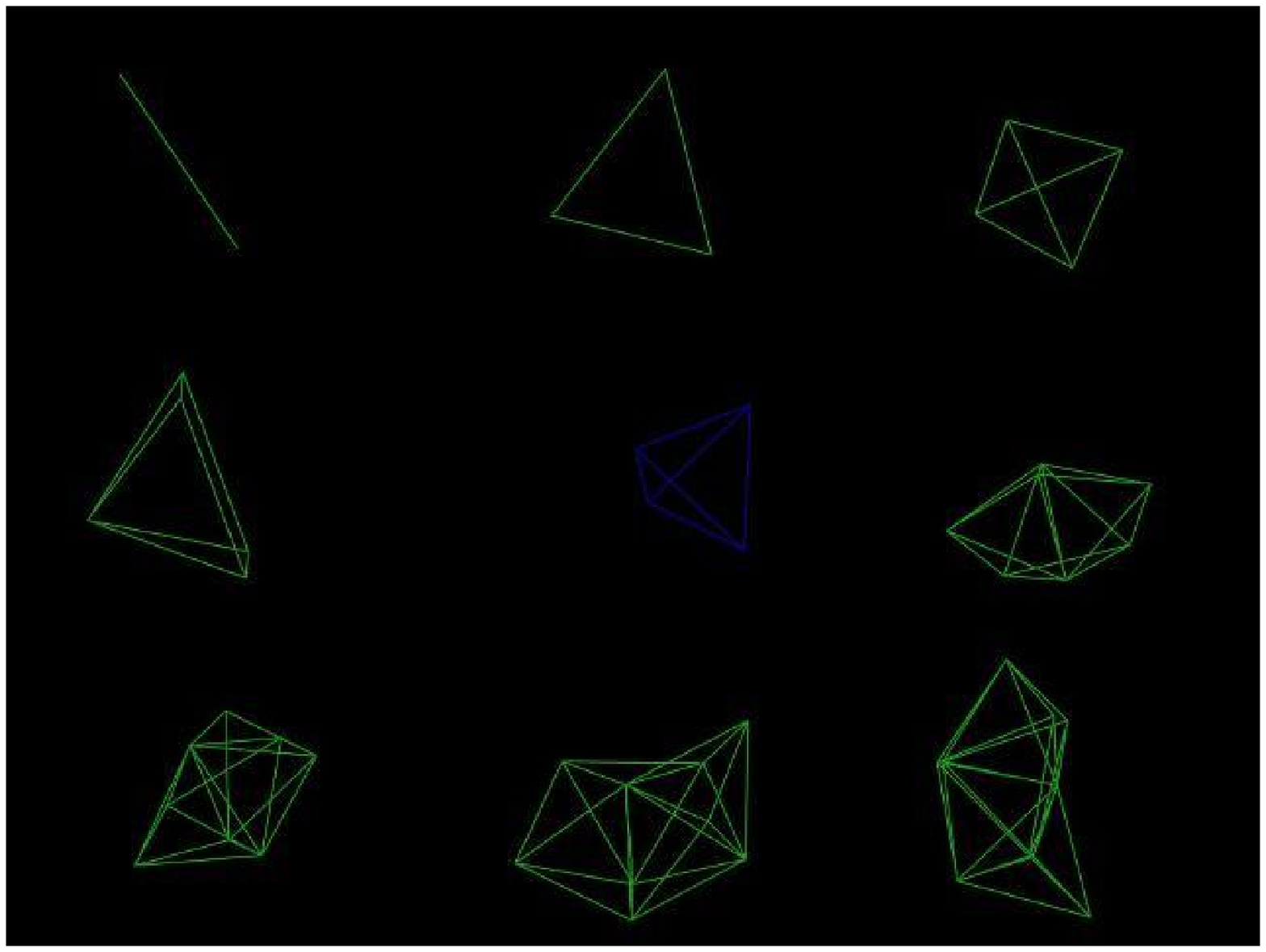}
\text{N=3, 4, 5, 6, 7, 8, 9, 10, 11 (\cite{web1,romerobg})}
\end{figure}

\begin{figure}
 \label{ouratoms}
 \centering
\includegraphics[scale=0.4]{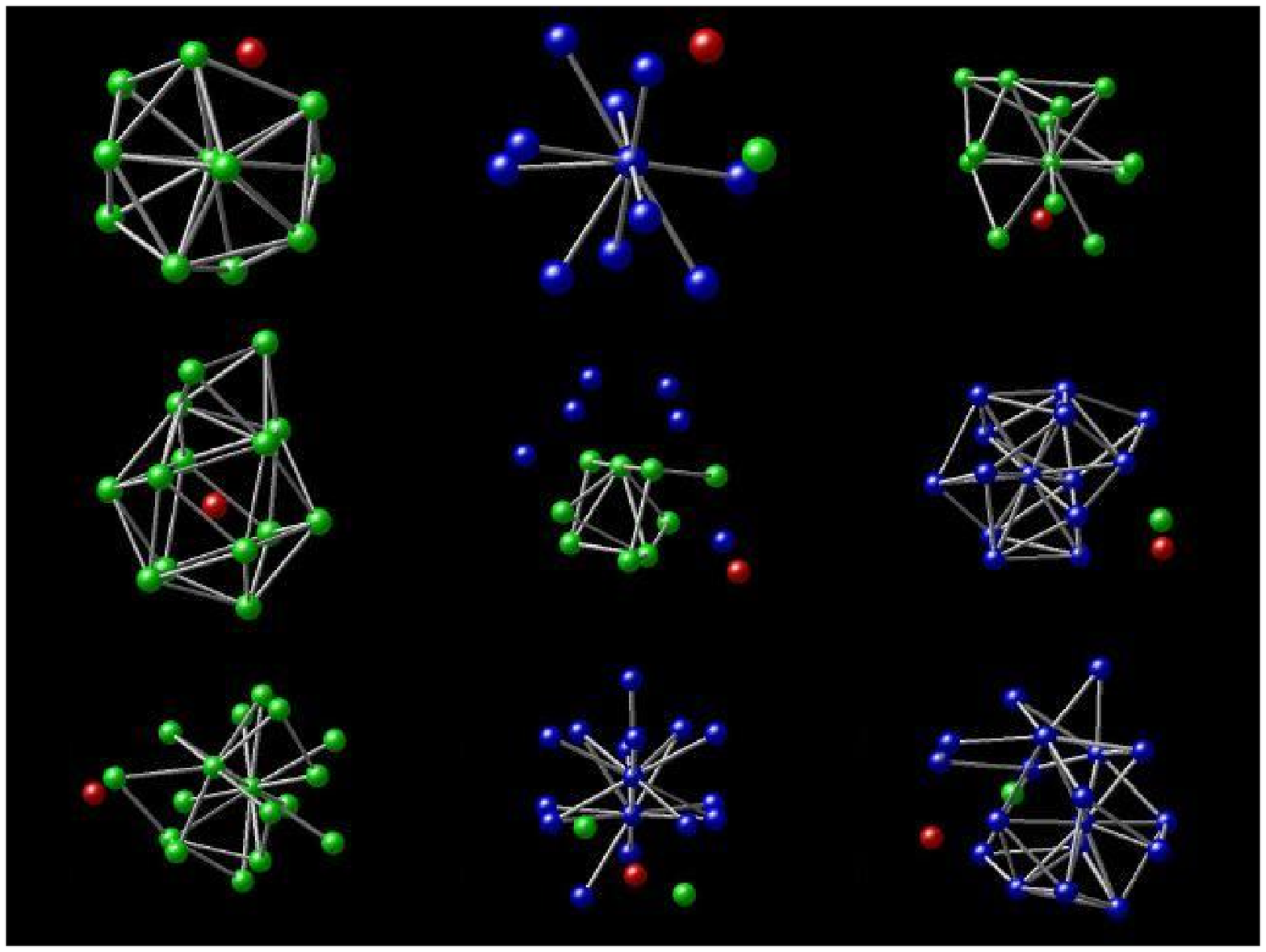}
\includegraphics[scale=0.4]{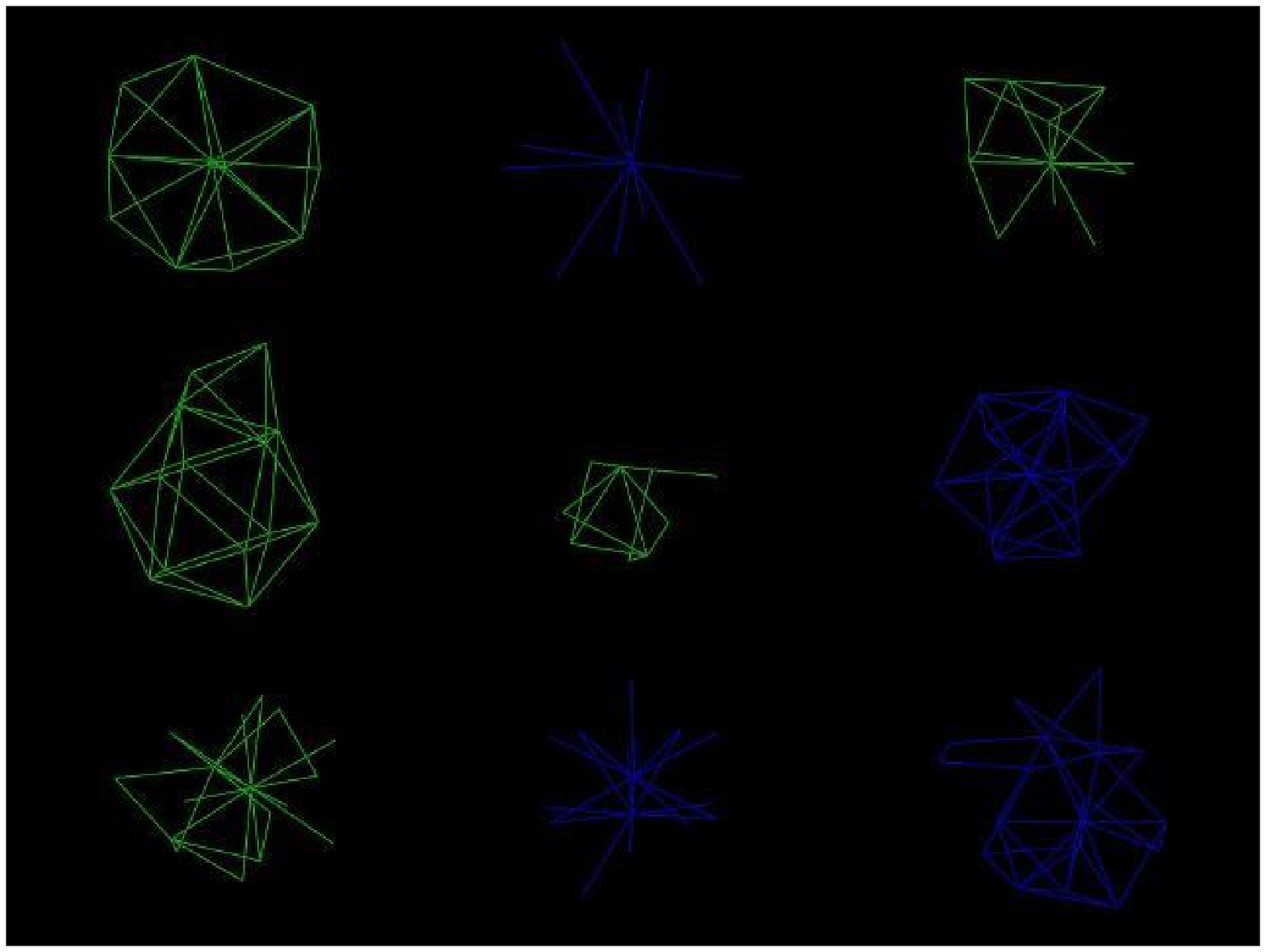}
\text{N=12, 13, 14, 15, 16, 17, 18, 19, 20 (ours)}
\end{figure}

\begin{figure}
 \label{ouratoms}
 \centering
\includegraphics[scale=0.4]{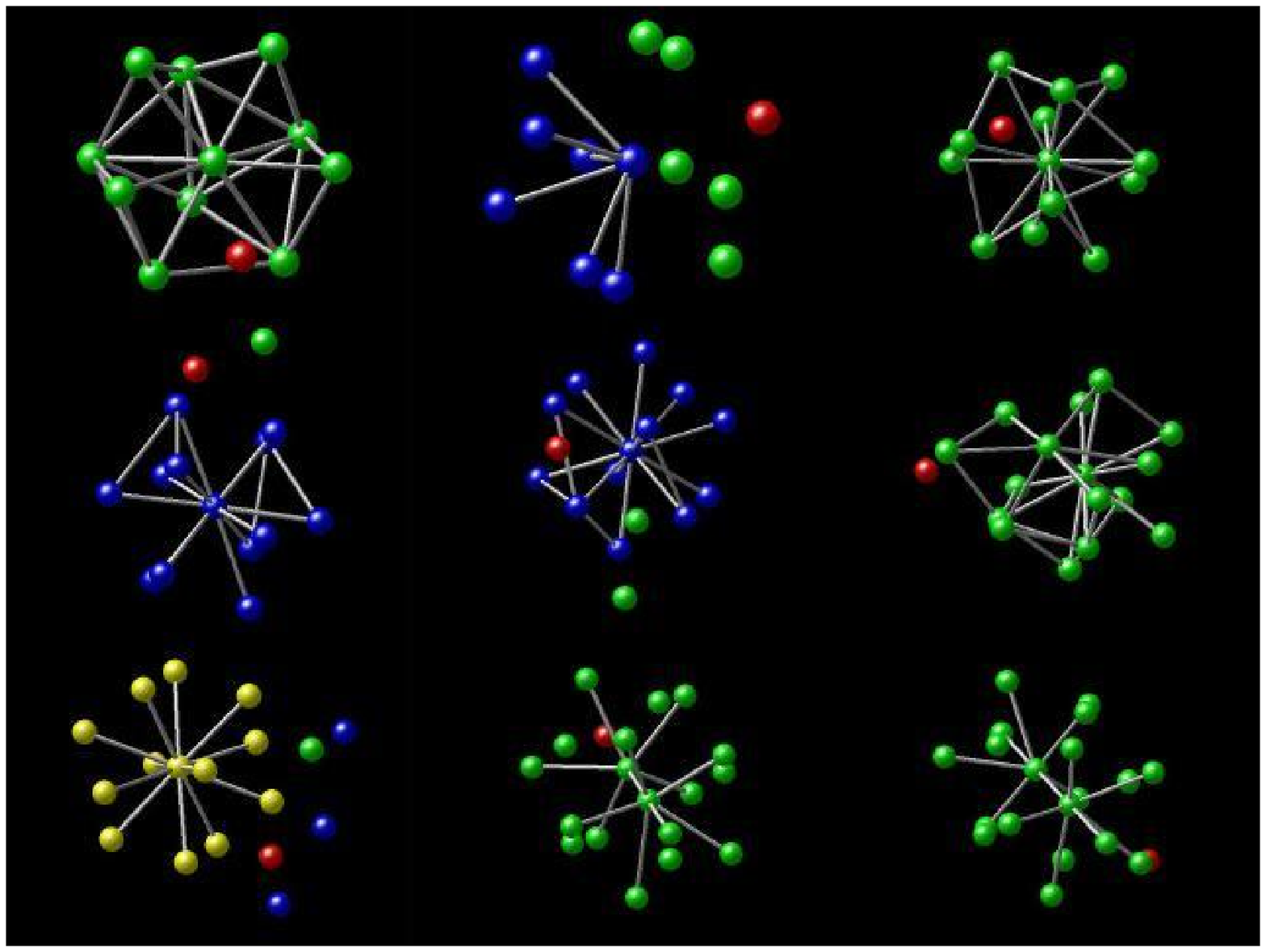}
\includegraphics[scale=0.4]{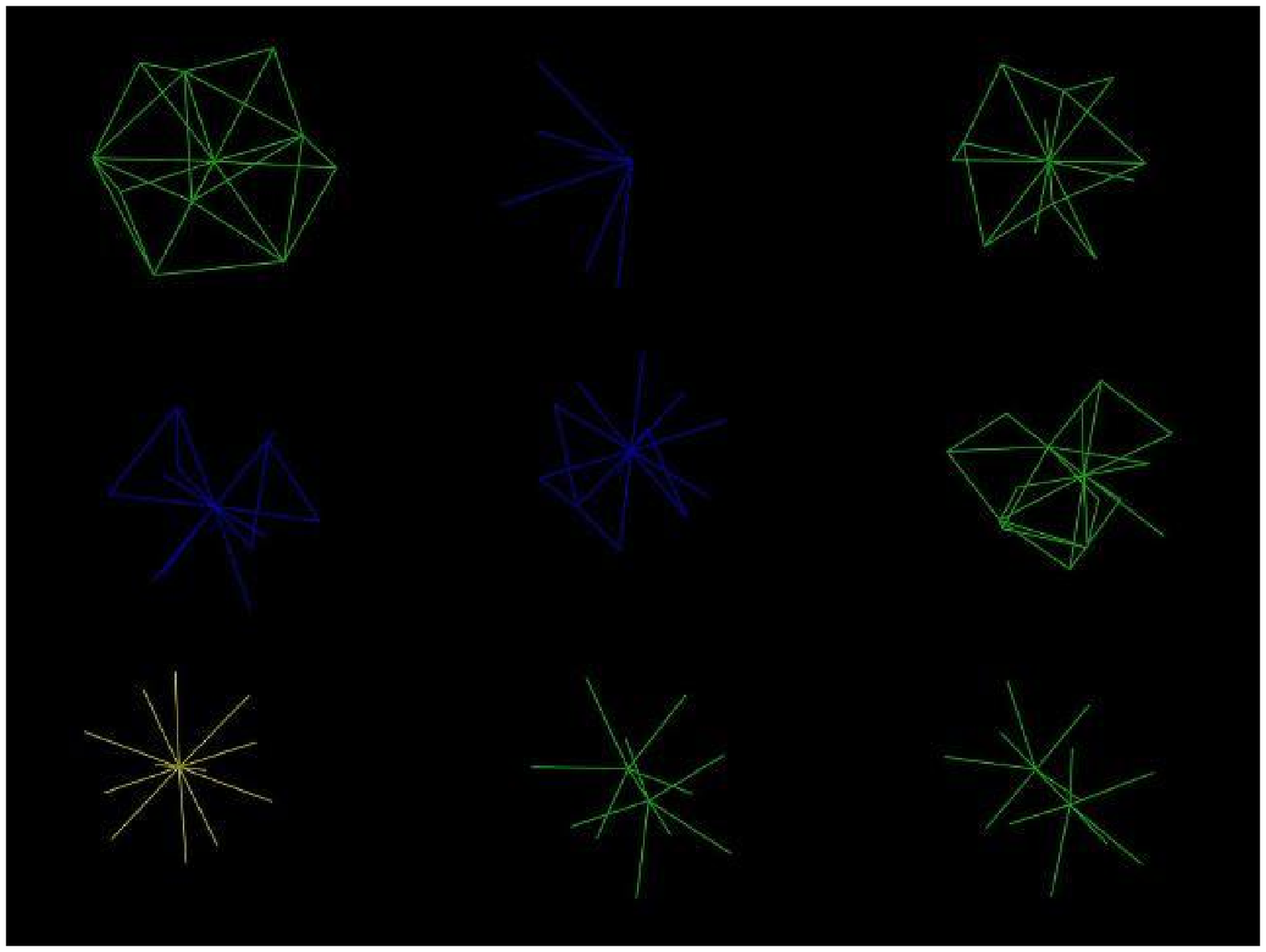}
\text{N=12, 13, 14, 15, 16, 17, 18, 19, 20 (\cite{web1,romerobg})}
\end{figure}

\begin{figure}
 \label{ouratoms}
 \centering
\includegraphics[scale=0.4]{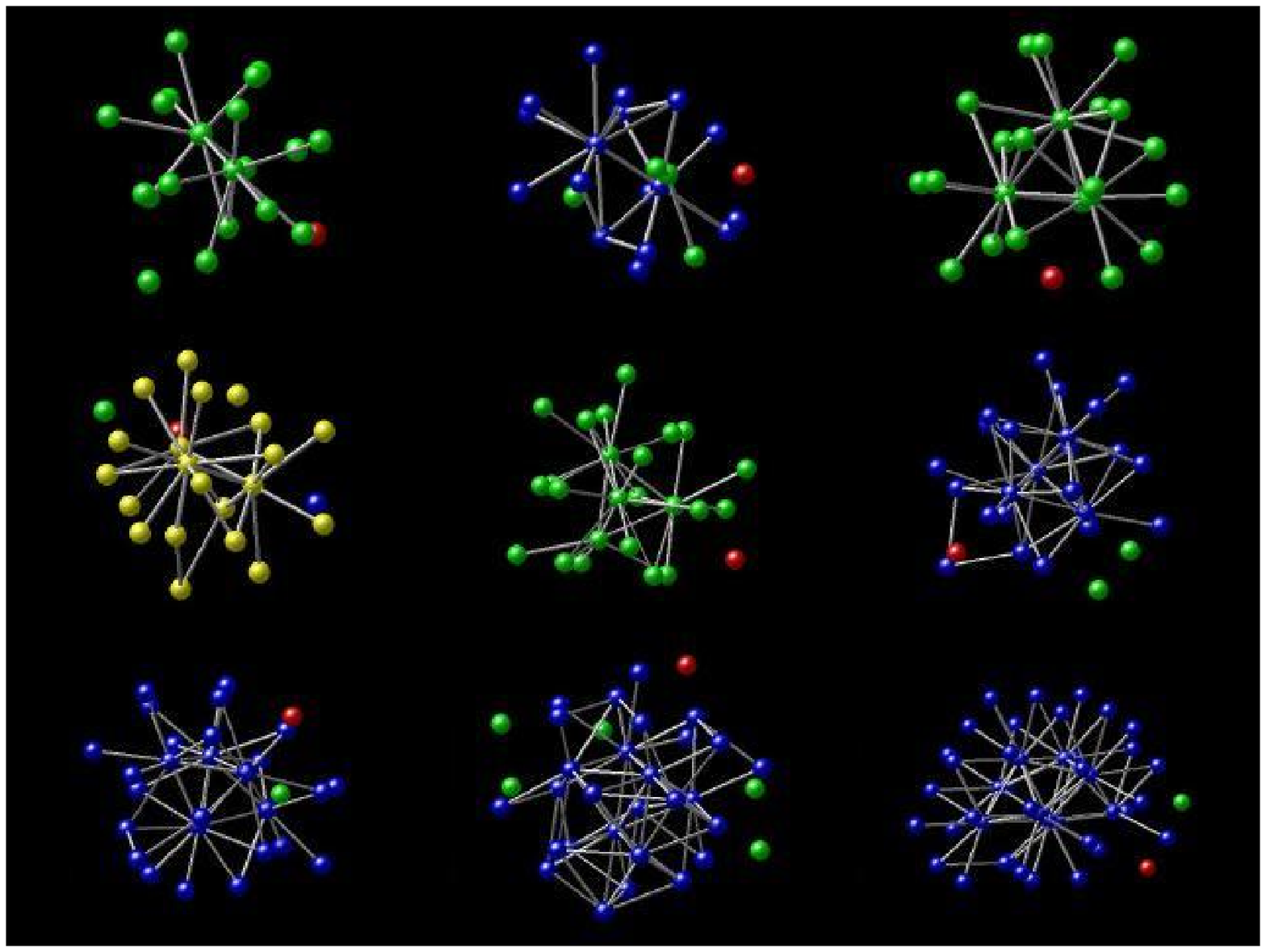}
\includegraphics[scale=0.4]{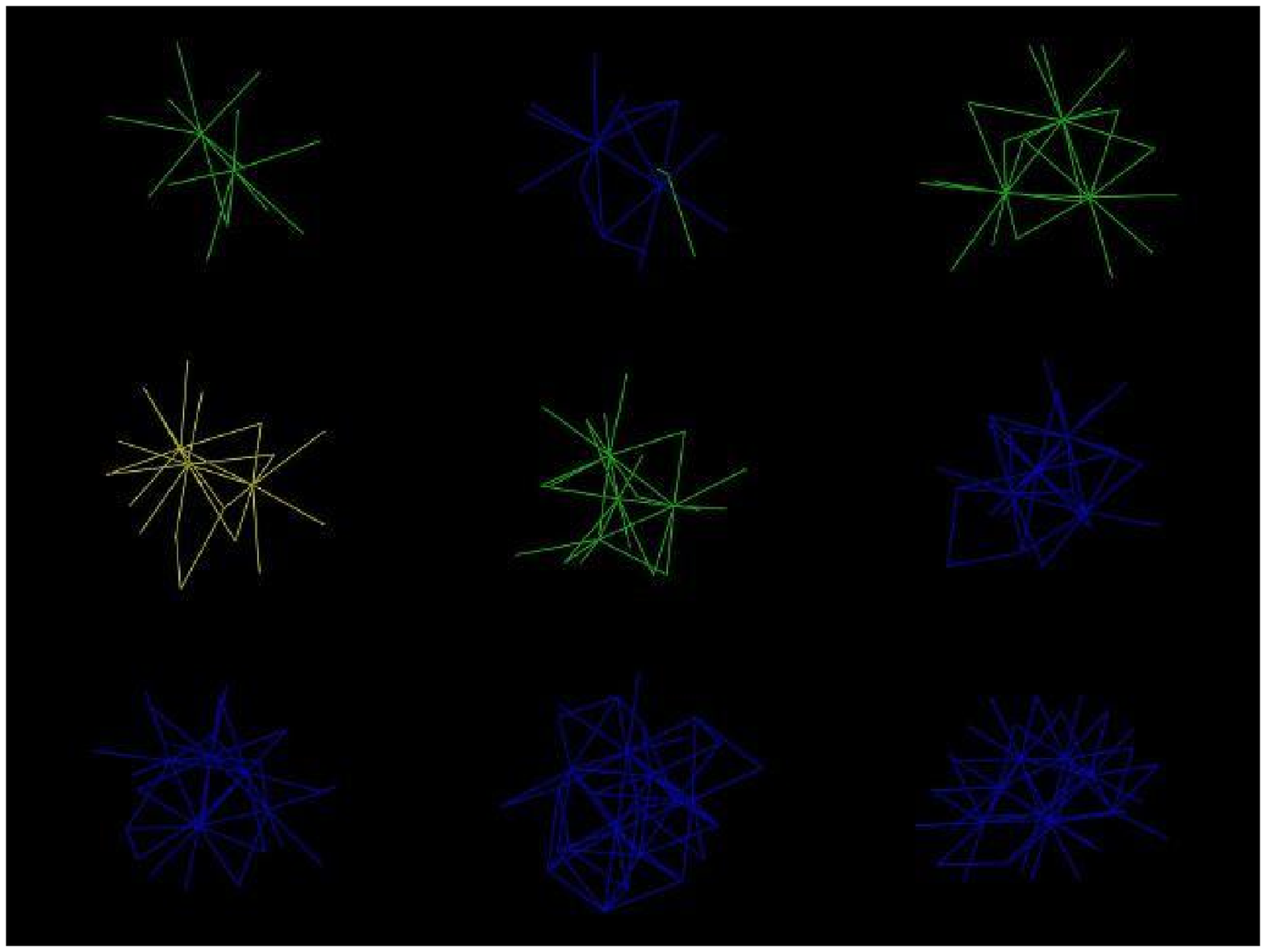}
\text{N=21, 25, 27, 30, 34, 44 (ours)}
\end{figure}

\begin{figure}
 \label{ouratoms}
 \centering
\includegraphics[scale=0.4]{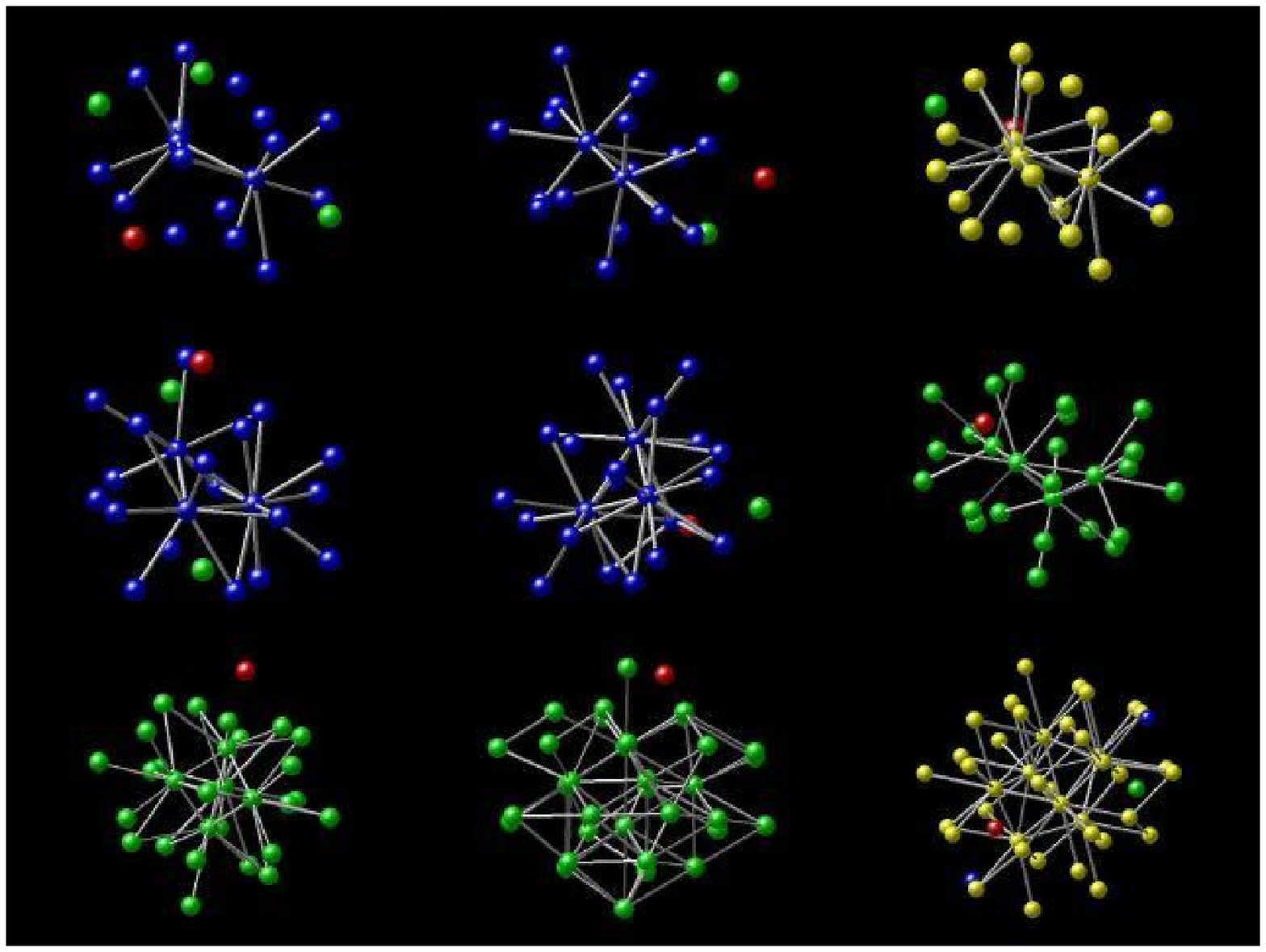}
\includegraphics[scale=0.4]{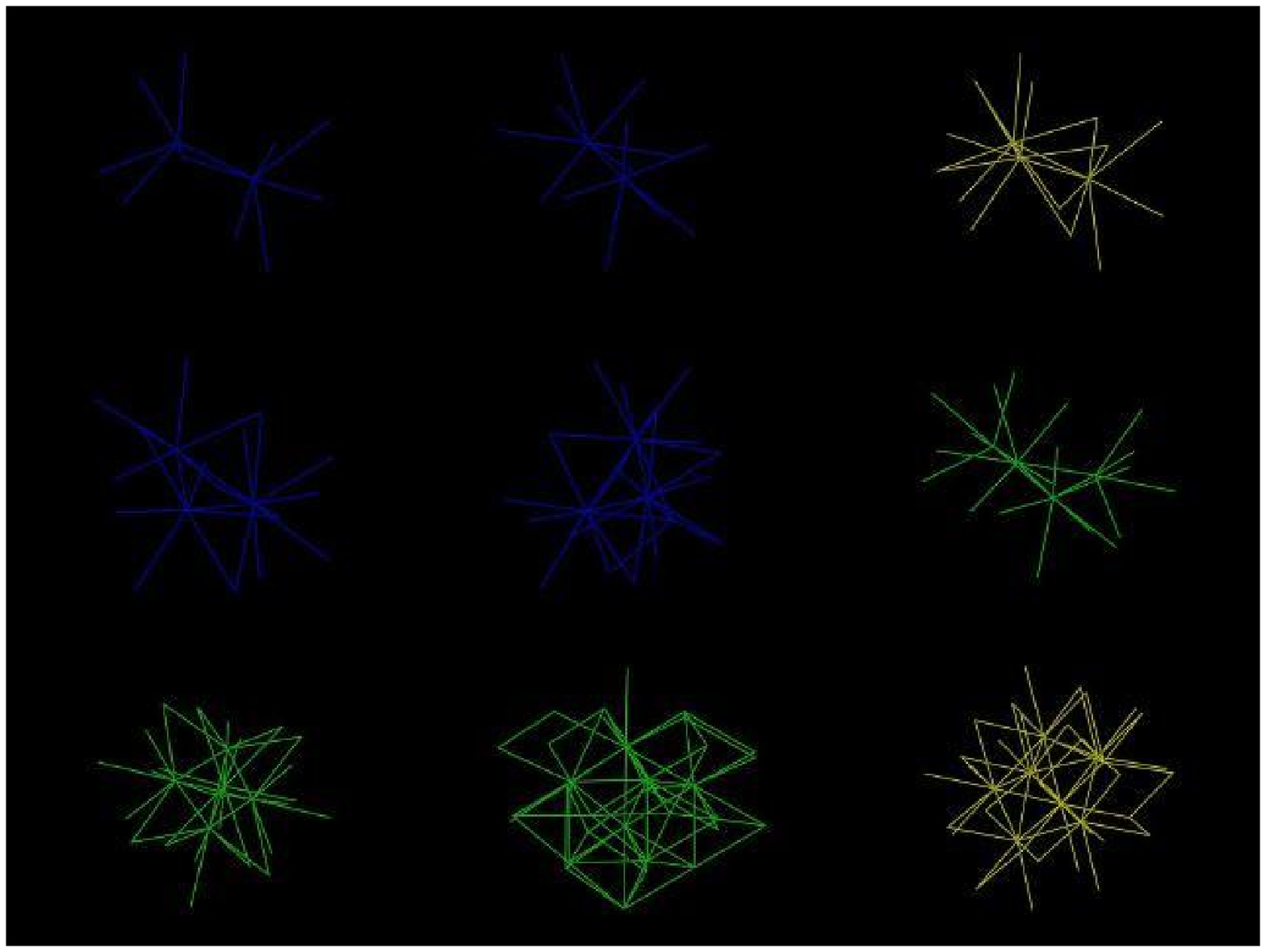}
\text{N=21, 25, 27, 30, 34, 44 (\cite{web1,romerobg})}
\end{figure}

\begin{figure}
 \label{ouratoms}
 \centering
\includegraphics[scale=0.4]{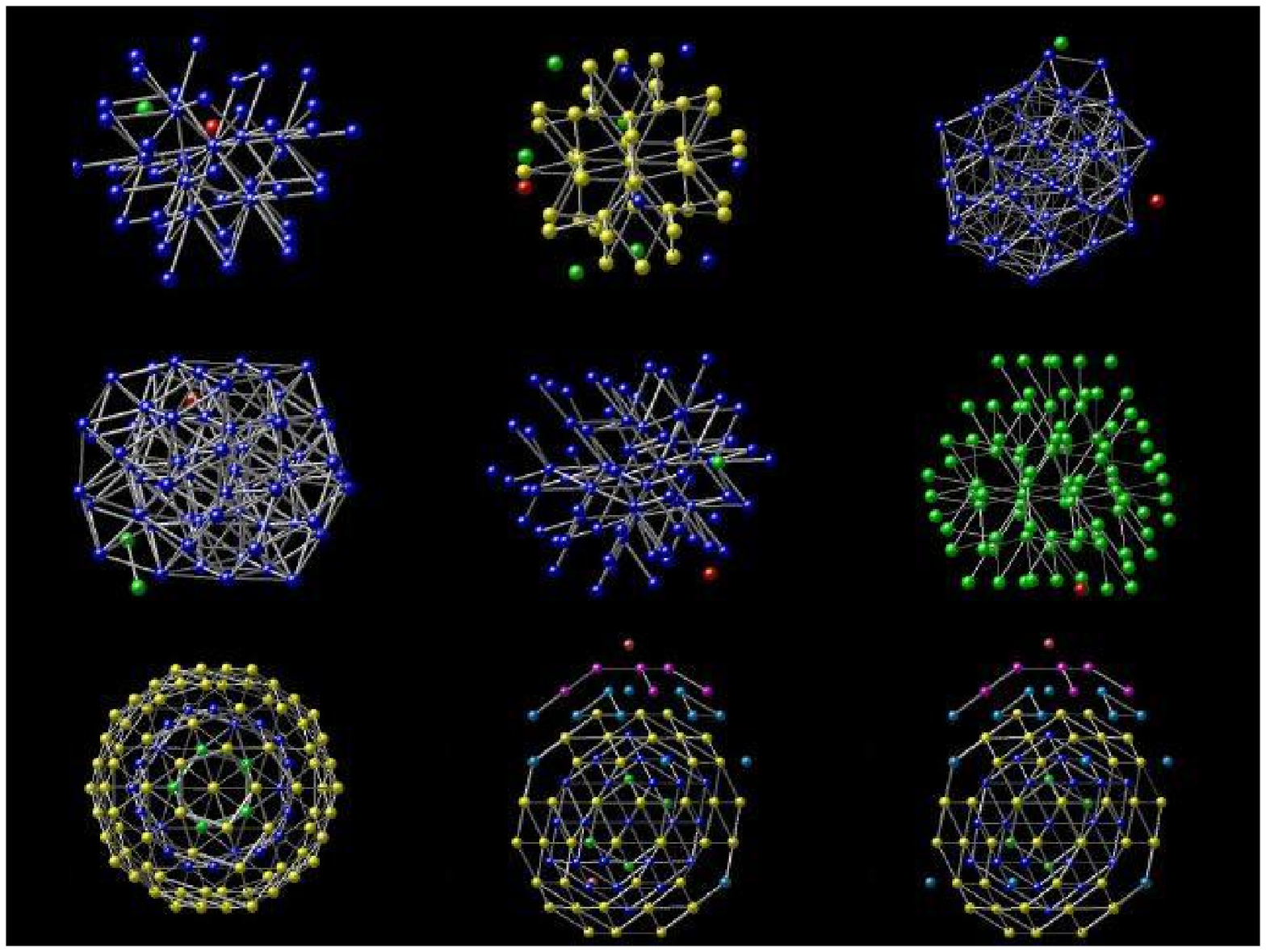}
\includegraphics[scale=0.4]{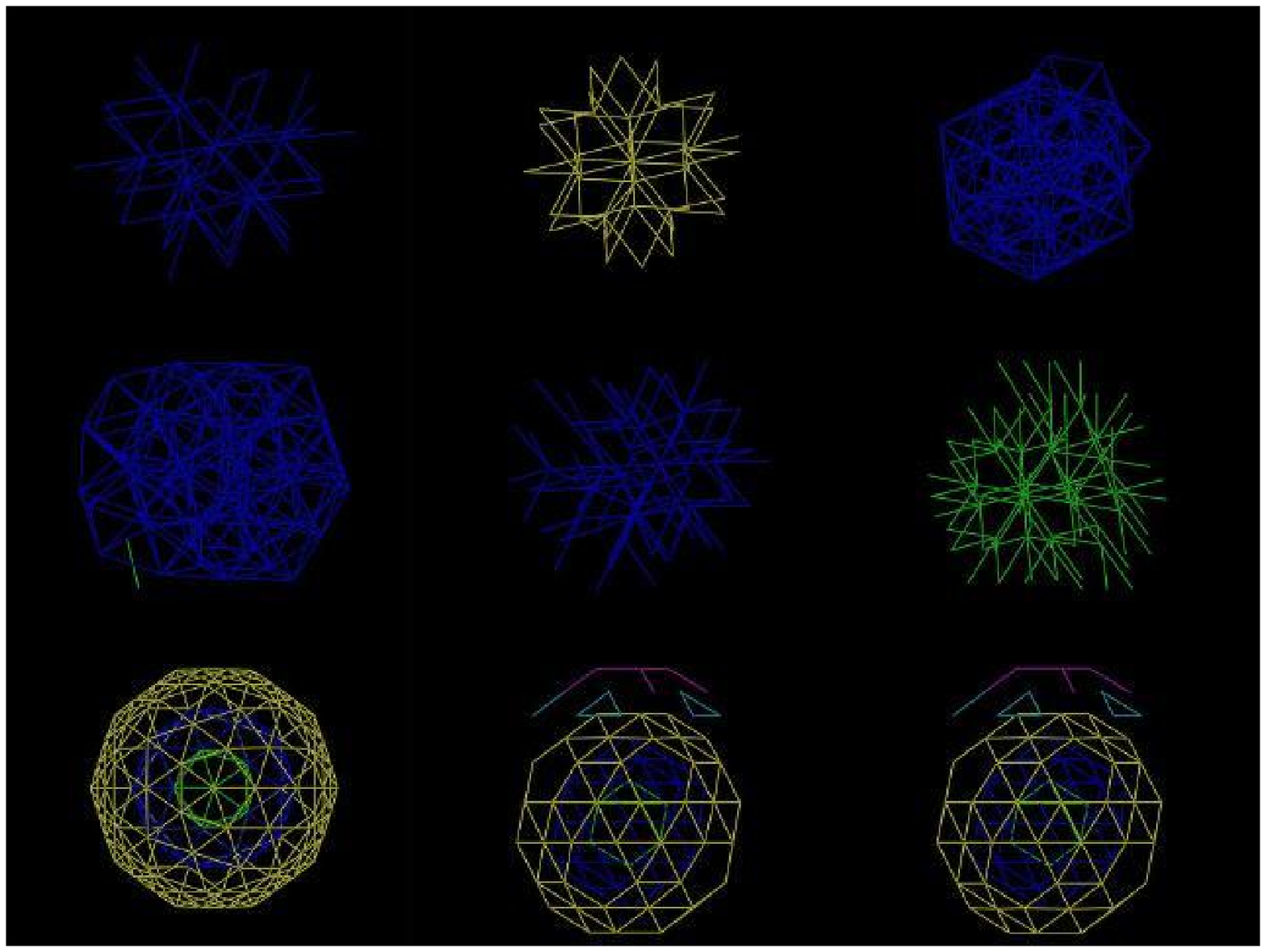}
\text{N=49, 56, 65, 67, 84, 93, 148, 170, 172 (ours)}
\end{figure}

\begin{figure}
 \label{ouratoms}
 \centering
\includegraphics[scale=0.4]{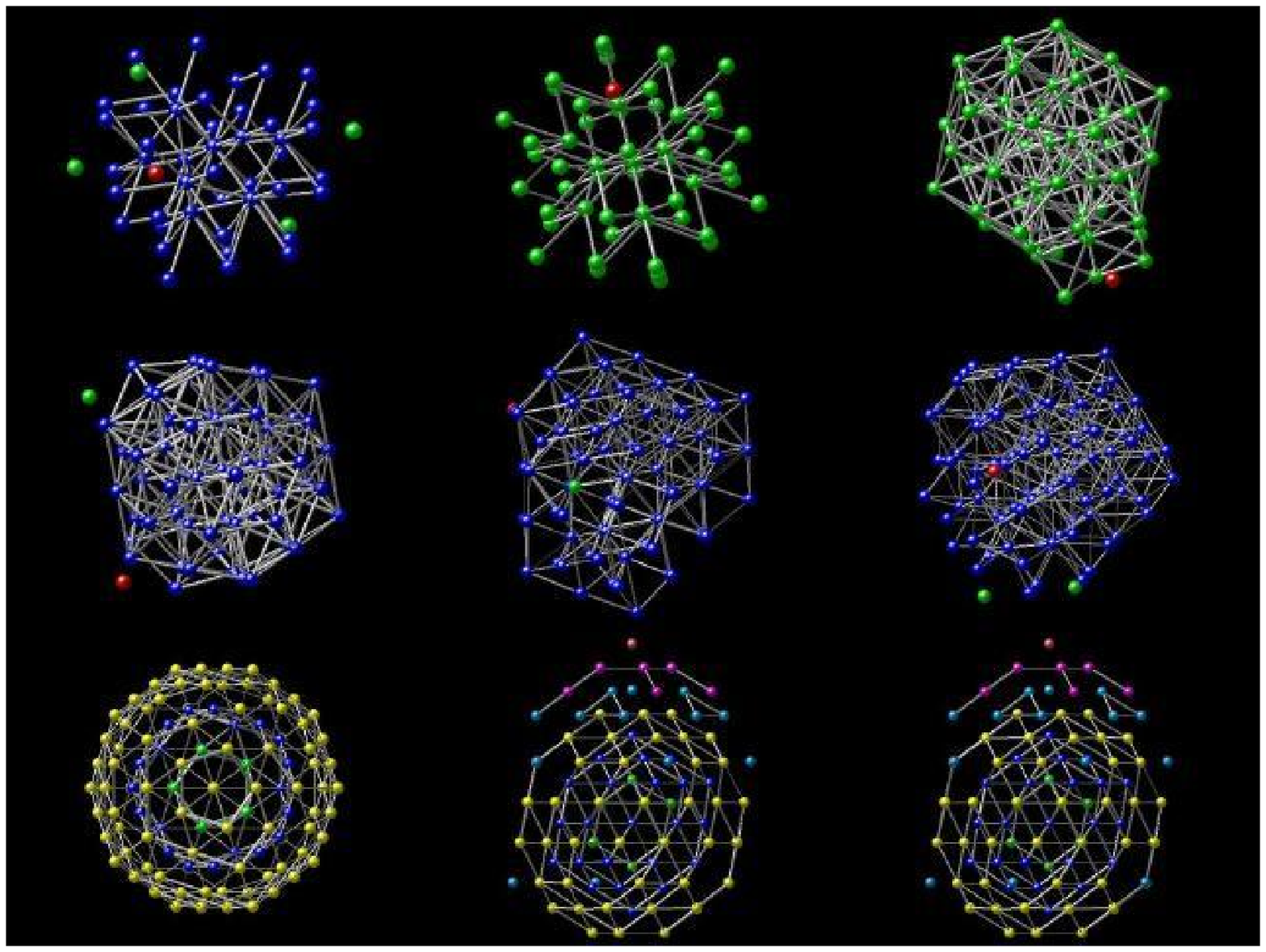}
\includegraphics[scale=0.4]{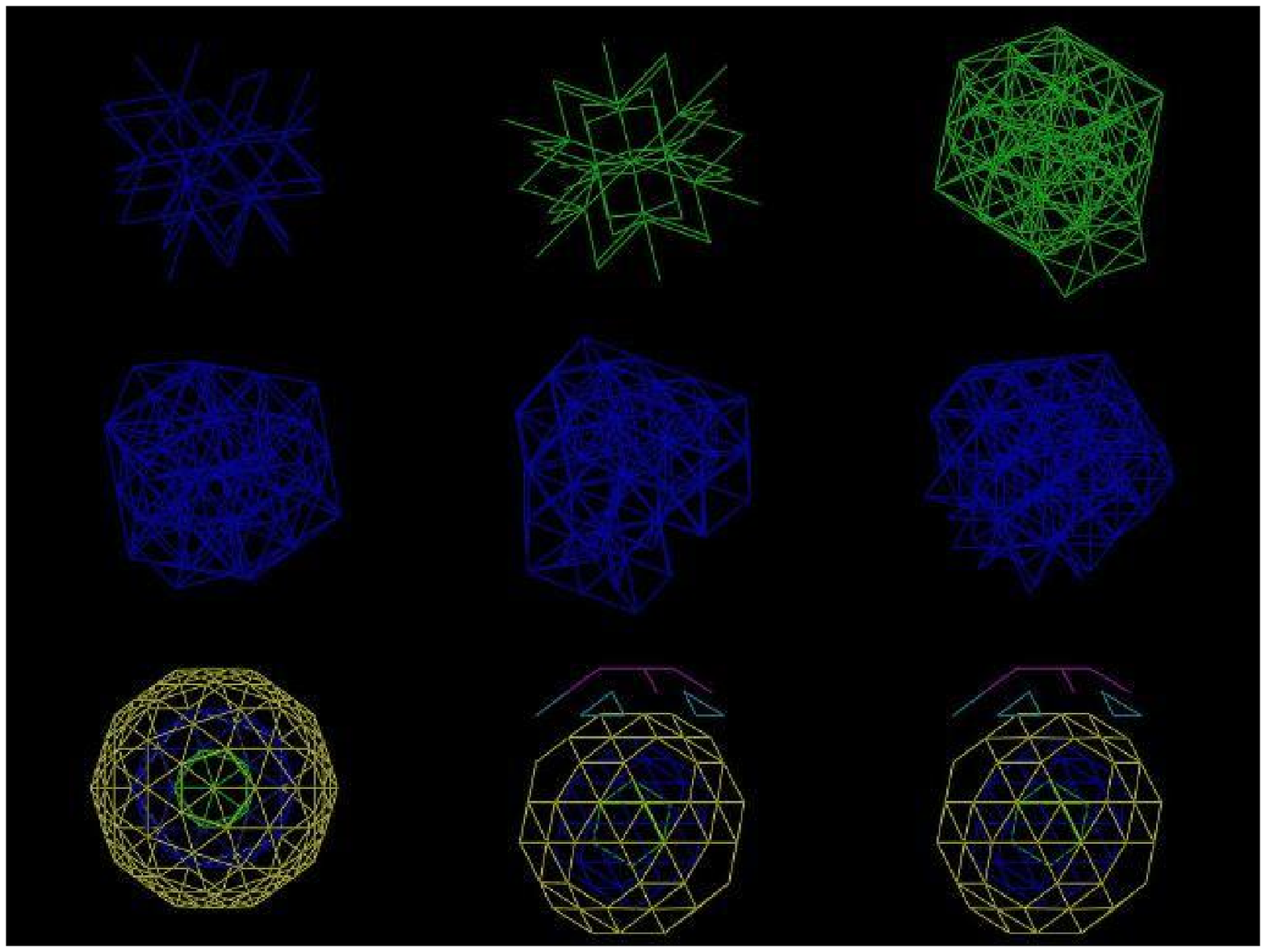}
\text{N=49, 56, 65, 67, 84, 93, 148, 170, 172
(\cite{web1,romerobg})}
\end{figure}

\begin{figure}
 \label{ouratoms}
 \centering
\includegraphics[scale=0.4]{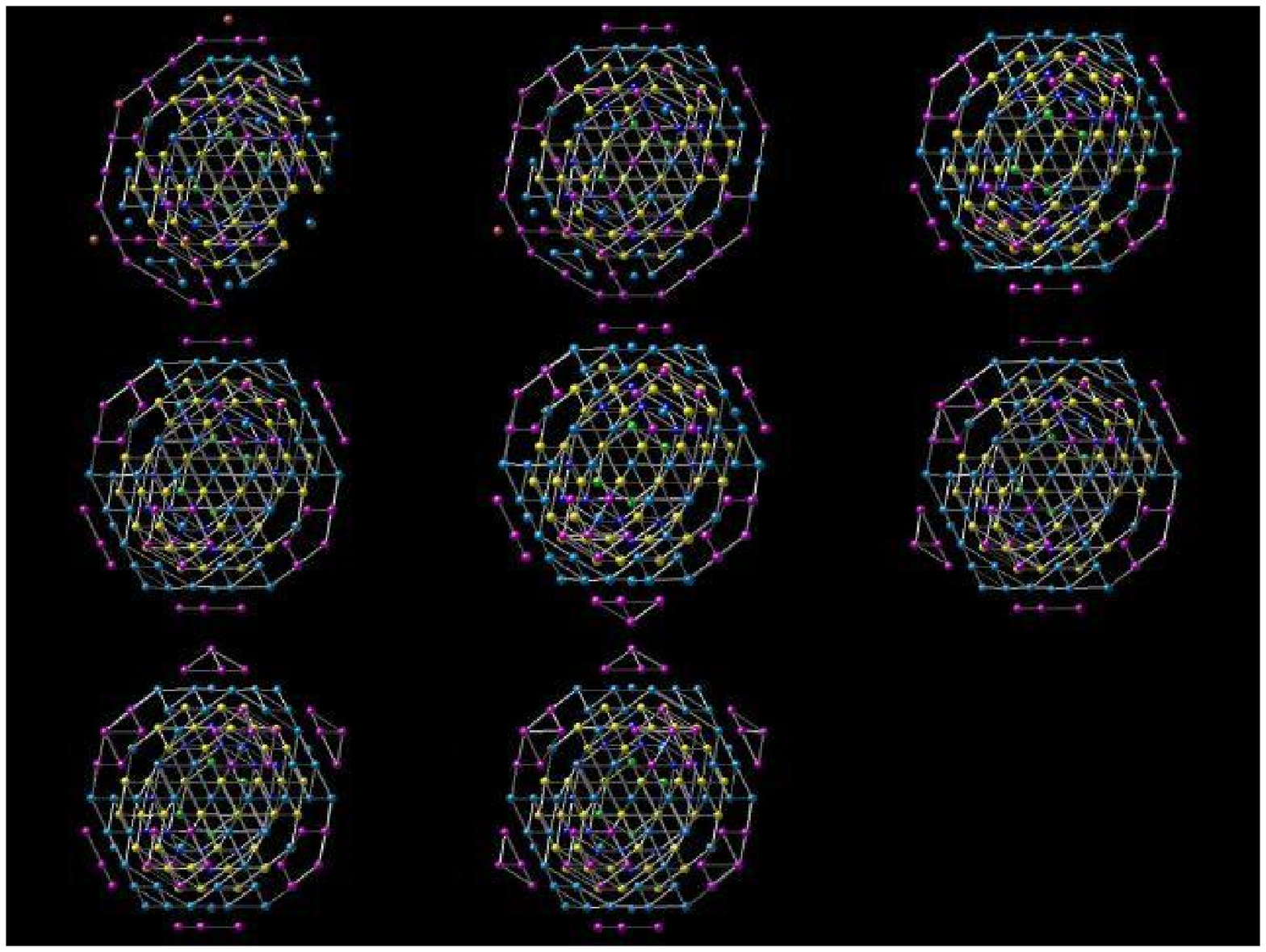}
\includegraphics[scale=0.4]{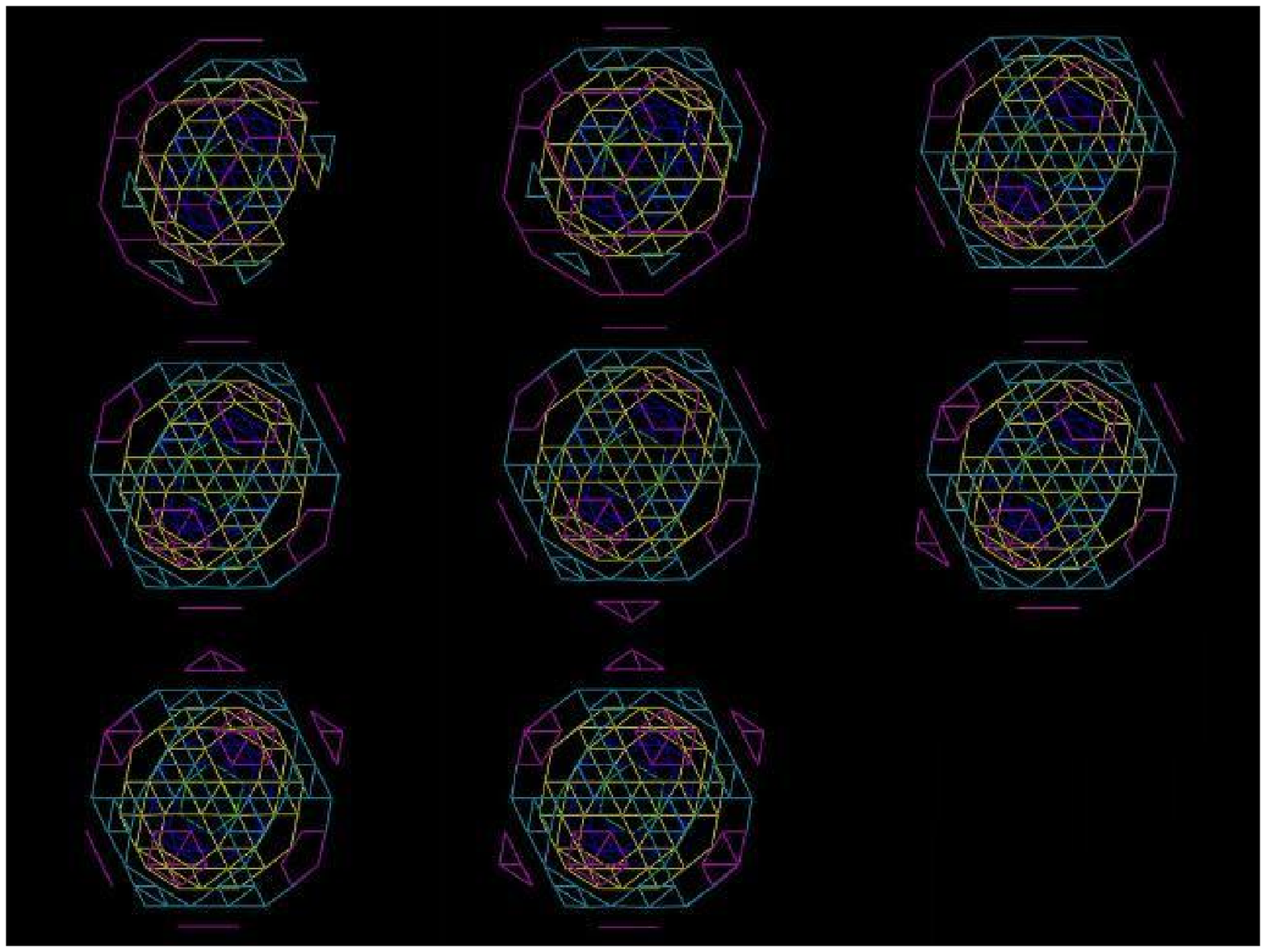}
\text{N=268, 288, 293, 298, 300, 301, 304, 308 (ours)}
\end{figure}

\begin{figure}
 \label{ouratoms}
 \centering
\includegraphics[scale=0.4]{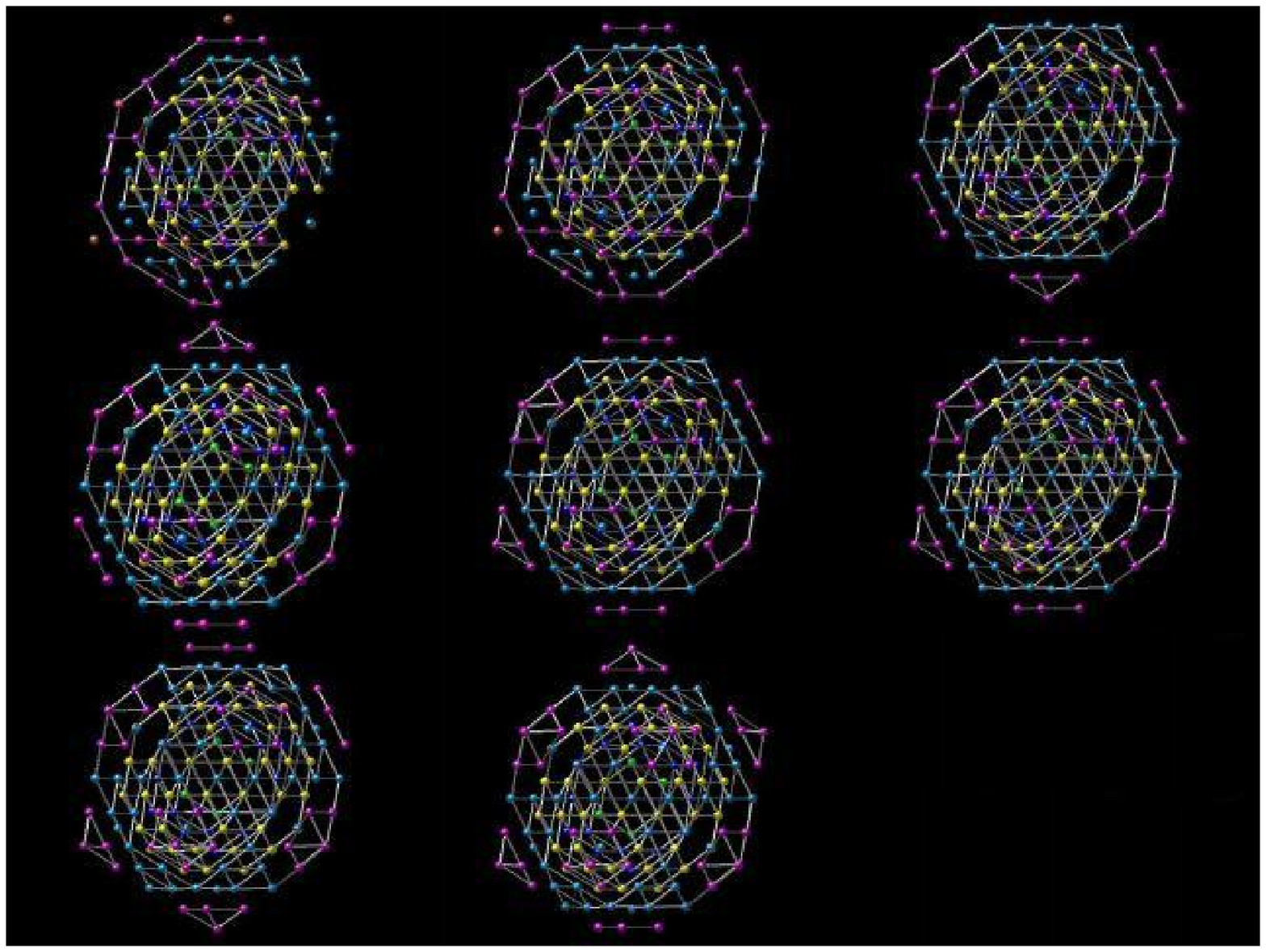}
\includegraphics[scale=0.4]{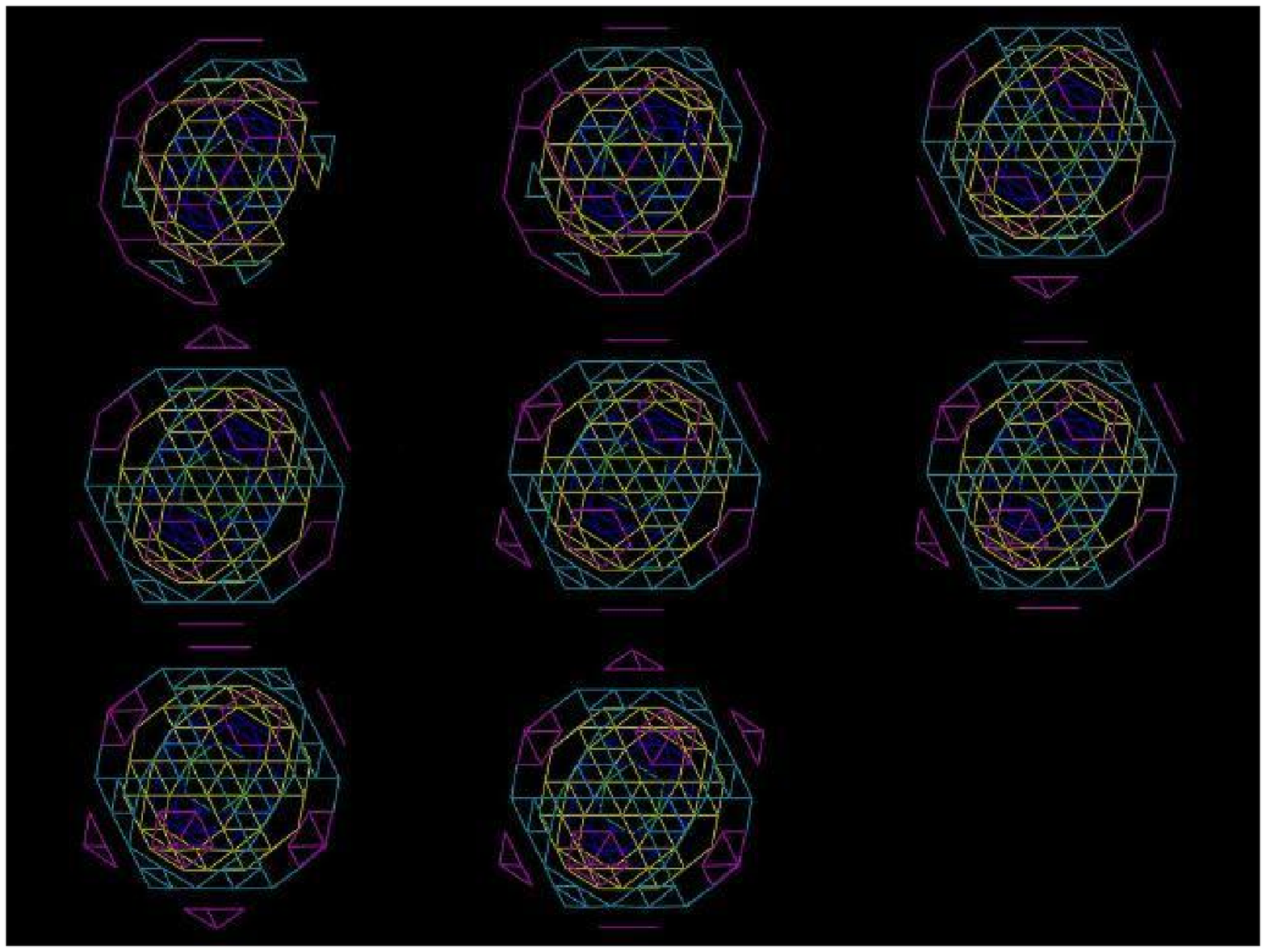}
\text{N=268, 288, 293, 298, 300, 301, 304, 308
(\cite{web1,romerobg})}
\end{figure}


\section{Conclusions} \label{concl}
In this paper, a brief review and a bibliography of important computational algorithms on minimizing the LJ potential energy are introduced in Sections 1 and 2. Section 3 of this paper illuminates many beautiful graphs for the three dimensional structures of molecules with minimal LJ potential. 

\begin{center}
{\bf Acknowledgements}
\end{center}
This research was supported by the Victorian Partnership for
Advanced Computing (http://www.vpac.org) of Australia.

\end{document}